\documentclass[journal=jctcce,manuscript=article]{achemso}
\usepackage{amssymb,amsmath,amsfonts,multicol,multirow,longtable,array,mathpazo}
\usepackage{lscape}
\usepackage[version=3]{mhchem} 
\usepackage{appendix}
\usepackage{soul} 
\usepackage[usenames]{color}
\usepackage{makecell}
\usepackage{enumerate}
\usepackage{xr}
\usepackage[T1]{fontenc} 
\usepackage{booktabs}
\usepackage{tikz}
\usepackage{threeparttable}
\usepackage{graphicx}
\usepackage{subfigure}
\usepackage{colortbl}
\usepackage{framed}
\usepackage{verbatim}
\usepackage{amsbsy}
\usepackage{bm}
\usepackage{bbm}
\usepackage[normalem]{ulem}
\usepackage{float}
\usepackage[linesnumbered,boxed]{algorithm2e}
\usepackage{algorithm2e}
\usepackage{amsmath}
\usepackage{diagbox}
\usepackage{simplewick} 

\newcounter{xscheme}

\newcommand{\sgn}{ {\rm{sgn}} }

\newfloat{Algorithm}{htbp}{alg}
\floatname{Algorithm}{Algorithm}

\newcounter{exe}[figure]
\newcommand{\iexe}{\refstepcounter{exe}\the\value{exe}:}

\setkeys{acs}{maxauthors = 0} 

\author{Wenjian Liu}\email{liuwj@sdu.edu.cn}
\affiliation{Qingdao Institute for Theoretical and Computational Sciences, Institute of Frontier and Interdisciplinary Science,
Shandong University, Qingdao, Shandong 266237, China}

\title{Perspective: ``Relativity + Correlation + QED = Experiment''}

\setkeys{acs}{articletitle=true}

\begin{document}

\begin{abstract}
The ultimate goal of electronic structure calculations is to make the left and right hand sides of the titled ``equation''
as close as possible. This requires high-precision treatment of relativistic, correlation, and quantum electrodynamics (QED)
effects simultaneously. While both relativistic and QED effects can readily be built into
the many-electron Hamiltonian, electron correlation is more difficult to describe due to the exponential growth
of the number of parameters in the wave function. Compared with the spin-free case, spin-orbit interaction results in
the loss of spin symmetry and concomitant complex algebra, thereby rendering the treatment of electron correlation
even more difficult. Possible solutions to these issues are highlighted here.
\end{abstract}

\maketitle


\section{Introduction}
Any electronic structure calculation ought to first choose an appropriate Hamiltonian (equation) and then an appropriate wave function Ansatz (method),
according to the target problem and accuracy. After a brief summary of relativistic Hamiltonians in Sec. \ref{SecH},
we discuss in Sec. \ref{SecEc} the construction of many-electron wave functions, focusing mainly on the static-dynamic-static (SDS) framework\cite{SDS}
for strongly correlated systems of electrons. Possible ways for a balanced and adaptive
treatment of electron correlation and spin-orbital coupling (SOC) are further highlighted therein. The paper ends up with perspectives
in Sec. \ref{Conclusion}.
\section{Relativistic Hamiltonians}\label{SecH}
\begin{figure}
\centering
\begin{center}
\includegraphics[width=0.5\columnwidth,keepaspectratio=true]{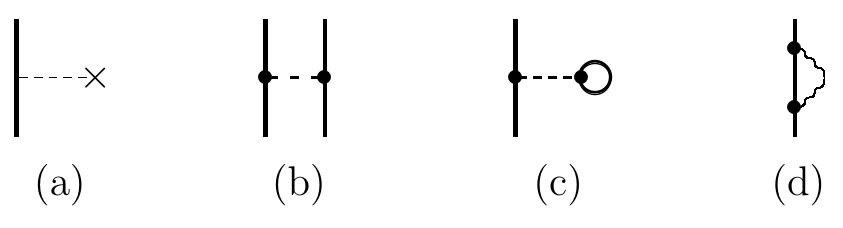}
\caption{Diagrammatical representation of $H_n^{\mathrm{eQED}}$ \eqref{HnH}: (a) one-body term;
(b) two-body term; (c) vacuum polarization \eqref{QVP}; (d) electron self-energy \eqref{QSE}.}
\label{Figure1}
\end{center}
\end{figure}

As the highest level of theory for electromagnetic interactions between charged particles (electrons, positrons, and nuclei),
quantum electrodynamics (QED) has achieved great success in ultrahigh-precision electronic structure calculations of few-body
systems (up to five electrons\cite{QEDeffects}). However, it can hardly be applied to
many-body systems due to its tremendous computational cost and complexity on the one hand, and its underlying philosophy on the other hand:
QED assumes from the outset that relativistic and QED effects are dominant over electron correlation, thereby following the
``first relativity and QED then correlation'' paradigm. It is obvious that this is only true for heavy ions of few electrons. Moreover,
the underlying time-dependent perturbation formalism is not suited to a high-order treatment of electron correlation: The more electrons and higher orders,
the more diagrams need to be included. It is clear that, for many-electron systems, it is relativity and correlation that should be
accounted for prior to QED effects, in a time-independent manner, thereby leading to the ``first relativity and correlation and then QED''
paradigm\cite{IJQCrelH}. This requires a many-body effective QED (eQED) Hamiltonian, $H_{\mathrm{eQED}}$, that is linear
in the electron-electron interaction, such that $H_{\mathrm{eQED}}\Psi=E\Psi$ is just a standard eigenvalue equation.
$H_{\mathrm{eQED}}$ is an effective Hamiltonian in the sense that it acts on the fermion Fock space as compared with
the much larger fermion-photon Fock space of QED. A natural route to obtain such a no-photon Hamiltonian is to extract appropriate operators from
the lowest-order QED energies\cite{np-eQED}. Interestingly, a completely bottom-up procedure (i.e.,
without recourse to QED at all) can also be invoked, either algebraically\cite{eQED}
or diagramatically\cite{IJQCeQED}. Briefly, the procedure starts with the famous filled
Dirac picture\cite{DiracSea} but incorporates charge conjugation (see, e.g., Ref. \citenum{eQEDBook2017}) in a proper way.
To see this, let us start with the following second-quantized Hamiltonian (under the Einstein summation convention),
\begin{eqnarray}
H&=&D_p^q a^p_q +\frac{1}{2}g_{pq}^{rs}a^{pq}_{rs},\quad p, q, r, s\in\mbox{ PES, NES},\label{Hbase}\\
D_p^q&=&\langle\psi_p|D|\psi_q\rangle,\quad g_{pq}^{rs}=\langle\psi_p\psi_q|g_{12}|\psi_r\psi_s\rangle,\label{Gpqrs}\\
a^p_q&=&a_p^\dag a_q,\quad a^{pq}_{rs}=a^\dag_pa^\dag_qa_sa_r,
\end{eqnarray}
which is normal ordered with respect to the genuine vacuum $|0\rangle$ of no particles nor holes
(PES: positive-energy states; NES: negative-energy states). Here,
$D$ consists of the Dirac operator and nuclear attraction, $g_{12}$ represents the electron-electron interaction,
and $\{\epsilon_p,\psi_p\}$ are eigenpairs of the mean-field operator $D+V_{\mathrm{eff}}(\boldsymbol{r})$.
The filled Dirac picture can then be realized in a finite basis representation by
setting the Fermi level below the energetically lowest of the $\tilde{N}$ occupied NES $\{\tilde{i}\}$. The physical energy of an $N$-electron state can be calculated\cite{PCCPNES} as
the difference between those of states $\Psi(N;\tilde{N})$ and $\Psi(0;\tilde{N})$,
\begin{eqnarray}
E&=&\langle \Psi(N;\tilde{N})|\mathcal{H}|\Psi(N;\tilde{N})\rangle - \langle \Psi(0;\tilde{N})|\mathcal{H}|\Psi(0;\tilde{N})\rangle,\label{EFStot}
\end{eqnarray}
provided that the charge-conjugation symmetry is incorporated properly.
To do so, we first shift the Fermi level just above the top of the NES. This amounts to normal ordering
the Hamiltonian $H$ \eqref{Hbase} with respect to $|0;\tilde{N}\rangle$, the non-interacting, zeroth-order term of $|\Psi(0;\tilde{N})\rangle$.
Here, the charge-conjugated contraction (CCC) of fermion operators\cite{eQED}, e.g.,
\begin{eqnarray}
\acontraction[0.5ex]{}{a^p}{}{a_q}a^pa_q&=&\langle 0;\tilde{N}|\frac{1}{2}[a^p, a_q]|0;\tilde{N}\rangle,\quad \quad p, q\in\mbox{ PES, NES}\label{CCC0}\\
&=& \frac{1}{2}\langle 0;\tilde{N}|a^{\tilde{p}}a_{\tilde{q}}|0;\tilde{N}\rangle|_{\epsilon_{\tilde{p}}<0, \epsilon_{\tilde{q}}<0}
   -\frac{1}{2}\langle 0;\tilde{N}|a_qa^p                    |0;\tilde{N}\rangle|_{\epsilon_p>0, \epsilon_q>0}\label{CCsym}\\
&=&-\frac{1}{2}\delta^p_q \sgn(\epsilon_q), \quad \forall p, q, \label{CCC}
\end{eqnarray}
must be invoked, so as to obtain
\begin{eqnarray}
H&=&H_n^{\mathrm{eQED}}+C_n,\\
H_n^{\mathrm{eQED}}&=&D_p^q\{a^p_q\}_n +\frac{1}{2}g_{pq}^{rs}\{a^{pq}_{rs}\}_n + Q_p^q\{a^p_q\}_n\label{HnH},\\
Q_p^q&=&\tilde{Q}_p^q+\bar{Q}_p^q=-\frac{1}{2}\bar{g}_{ps}^{qs}\sgn(\epsilon_s),\label{QDef}\\
\tilde{Q}_p^q&=&-\frac{1}{2}g_{ps}^{qs}\sgn(\epsilon_s),\label{QVP}\\
\bar{Q}_p^q &=& \frac{1}{2}g_{ps}^{s q}\sgn(\epsilon_s)\label{QSE}.
\end{eqnarray}
The constant, zero-body term $C_n=\langle 0;\tilde{N}|H|0;\tilde{N}\rangle$ does not contribute to the physical energy
and will be `renomalized away'. Noticeably, the two effective one-body terms,
 $\tilde{Q}$  and $\bar{Q}$,
involve summations over all positive- and negative-energy states, a direct consequence of
the CCC \eqref{CCC0} (which treats
the filled negative-energy electron and positron seas on an equal footing, as it should be). Had the standard contraction of fermion operators,
$\langle 0;\tilde{N}|a^pa_q|0;\tilde{N}\rangle=\delta^{\tilde{p}}_{\tilde{q}}n_{\tilde{q}}$,
been taken, an infinitely repulsive potential, $\bar{g}_{p\tilde{i}}^{q\tilde{i}}n_{\tilde{i}}$, would be
obtained, such that no atom would be stable! As can be seen from Fig. \ref{Figure1}(c) and \ref{Figure1}(d),
the direct term $\tilde{Q}$ \eqref{QVP} and exchange term $\bar{Q}$ \eqref{QSE} are precisely the
vacuum polarization (VP) and electron self-energy (ESE), respectively. It also deserves to be mentioned that,
although the diagrams Fig. \ref{Figure1}(c) and \ref{Figure1}(d) are asymmetric, their weight factors are still $1/2$ instead of 1,
again due to the averaging of the negative-energy electron and positron seas.

The expression \eqref{HnH} can be rewritten in a more familiar form known from nonrelativistic quantum mechanics,
by further normal ordering with respect to $|N;\tilde{N}\rangle$, the non-interacting, zeroth-order term of $|\Psi(N;\tilde{N})\rangle$.
Since only PES
are involved here, the standard contraction of fermion operators, e.g.,
\begin{eqnarray}
\bcontraction[0.5ex]{}{a^p}{}{a_q}a^pa_q=\langle N;\tilde{N}|\{a^pa_q\}_n|N;\tilde{N}\rangle=\langle N;0|a^pa_q|N;0\rangle=\delta^p_q n_q, \quad \epsilon_q>0, \label{NRCCC}
\end{eqnarray}
should be invoked, thereby leading to
\begin{eqnarray}
H_n^{\mathrm{QED}}&=& E^{\mathrm{QED}}_{\mathrm{ref}}+f^{\mathrm{QED}}_{pq}\{a^p_q\}_F + \frac{1}{2}g_{pq}^{rs}\{a^{pq}_{rs}\}_F,\label{HeQED}\\
f^{\mathrm{QED}}_{pq}&=&f^{\mathrm{4C}}_{pq}+Q_p^q,\label{full-Fock}\\
f^{\mathrm{4C}}_{pq}&=&D_p^q+(V_{HF})_p^q,\quad (V_{HF})_p^q=\bar{g}_{pj}^{qj},\label{Feop}\\
E^{\mathrm{QED}}_{\mathrm{ref}}&=&\langle N;\tilde{N}|H_n^{\mathrm{eQED}}|N;\tilde{N}\rangle=E^{\mathrm{4C}}_{\mathrm{ref}}+Q_i^i,\label{E1full}\\
E^{\mathrm{4C}}_{\mathrm{ref}}&=&(D+\frac{1}{2}V_{HF})_i^i.\label{E1np}
\end{eqnarray}

\begin{figure}
\centering
\begin{center}
\includegraphics[width=0.5\columnwidth,keepaspectratio=true]{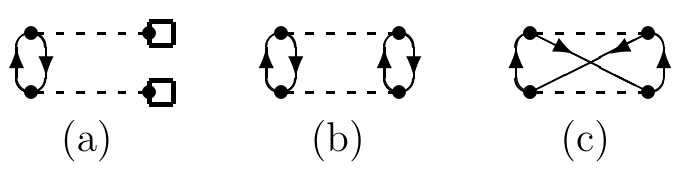}
\caption{Goldstone diagrams for the one-body (a) and two-body (b, c) terms of $E^{(2)}_{\mathrm{eQED}}$
($=E^{(2)}_{++}+E^{(2)}_{+-}$) as difference
between the two terms in Eq. \eqref{EFStot}. For the first term, the particles
(up-going lines) and holes (down-going lines) are $\{a, b\}$ and $\{i,j, \tilde{i}, \tilde{j}\}$,
respectively, whereas for the second term, they are $\{a, b, i, j\}$ and $\{\tilde{i}, \tilde{j}\}$,
respectively. Here, $\{p\}$ and $\{\tilde{p}\}$ refer to PES and NES, respectively.}
\label{FigureE2}
\end{center}
\end{figure}

For more comprehensive elucidations of the above bottom-up construction\cite{eQED,IJQCeQED} of $H_n^{\mathrm{eQED}}$ \eqref{HnH}/\eqref{HeQED},
we refer the reader to Refs. \citenum{IJQCrelH,IJQCeQED,PhysRep,X2C2016,Essential2020,RelChina2020}. Some brief remarks are sufficient here.
$H_n^{\mathrm{eQED}}$ differs from the eQED Hamiltonian obtained in a top-down fashion\cite{np-eQED} primarily in that
the latter adopts the no-pair approximation (NPA) from the outset (via projection operators that have to be chosen carefully), but which
is not invoked in $H_n^{\mathrm{eQED}}$.
As such, $H_n^{\mathrm{eQED}}$ \eqref{HnH}/\eqref{HeQED} represents the most accurate many-electron relativistic Hamiltonian,
and hence serves as the basis of ``molecular QED'' and new physics beyond the standard model of physics (see
recent investigations\cite{SaueeQED2022,ChengNewPhys2022} in this context, although QED effects were treated only approximately therein).
Given its short-range nature, the $Q$ potential \eqref{QDef} can readily be fitted into a model operator for each atom\cite{ShabaevModelSE,ShabaevModelSEcode2018}, so as to treat the VP-ESE (Lamb shift) variationally at the mean-field level.
Overall, the present time-independent formulation is not only elegant but also simplifies greatly the derivation of QED energies:
The full, M{\o}ller-Plesset (MP)-like second-order QED energy involves only three Goldstone diagrams in the present case
(cf. Fig. \ref{FigureE2}), but involves in total 28 Feynman diagrams
in the S-matrix formulation of QED\cite{eQED}.

Under the NPA, the four-component (4C) relativistic Hamiltonian, with or without the $Q$ term \eqref{QDef}, amounts to just retaining
the PES in $H_n^{\mathrm{eQED}}$ \eqref{HeQED}, viz.,
\begin{eqnarray}
H_+^{\mathrm{X}}=E^{\mathrm{X}}_{\mathrm{ref}}+f^{\mathrm{X}}_{pq}\{a^p_q\}_F + \frac{1}{2}g_{pq}^{rs}\{a^{pq}_{rs}\}_F,\quad p,q,r,s\in \mbox{ PES}; \quad \mathrm{X = eQED, 4C}.\label{H+4C}
\end{eqnarray}
The DHF operator \eqref{Feop} represented in a restricted kinetically balanced basis\cite{RKB}
can be block-diagonalzied in one step\cite{NESC,X2C2005,X2C2009},
so as to lead to the so-called exact two-component (X2C)\cite{X2Cname} approach,
\begin{eqnarray}
H_+^{\mathrm{X2C}}=E^{\mathrm{X2C}}_{\mathrm{ref}}+f^{\mathrm{X2C}}_{pq}\{a^p_q\}_F + \frac{1}{2}g_{pq}^{rs}\{a^{pq}_{rs}\}_F,\label{H+X2C}
\end{eqnarray}
where the orbitals now refer to the X2C two-component (2C) spinors.
Since picture-change effects\cite{PCE}
stem solely from the innermost core, they can readily be incorporated into the X2C Fockian $\mathbf{f}^{\mathrm{X2C}}$
(see equation (191) in Ref. \citenum{Essential2020}), thereby
leaving the two-body term untransformed.
For more details, see comprehensive  reviews\cite{PhysRep,X2C2016,Essential2020,RelChina2020,LiuMP,SaueRev,ReiherRev,X2CBook2017}.
In particular, the relationships between one- and multiple-step X2C's have been made crystal clear\cite{LiuMP}:
They share the same decoupling condition and differ only in the renormalization, and can all be
formulated either at matrix or operator level.
Due to its simplicity, accuracy, and efficiency\cite{Q4C,Q4CX2C,SaueX2C,DLU,X2Cgrd2020,SOX2CAMF}, the one-step
X2C\cite{X2C2005,X2C2009} has
become the workhorse of relativistic quantum chemical calculations under the NPA, as symbolized by its realization in a large number of
software packages\cite{BDF1,BDFrev2020,Dirac2020,COLOGNE15,Molcas8-2016,QCMAQUIS,PySCF2018,ResPect2020,
Psi42020,Turbomole2020,NWChem2020,CFOUR2020,Molpro2020,Chronus2020,Hyperion2022,ADF-X2C,Q-Chem-X2C2021}. Note that
the $Q$ term \eqref{QDef} can also be included in the transformation, so as to obtain eQED@X2C.

Albeit defined only algebraically, the X2C Hamiltonian can still be separated into a spin-free part and a spin-dependent part by means of matrix perturbation theory\cite{X2CSOC1,X2CSOC2,X2CSOCBook2017}, viz.,
\begin{eqnarray}
H^{\mathrm{X2CSOC}}_{+}&=&E_{\mathrm{ref}}^{\mathrm{X2CSOC}}+H_{\mathrm{sf}}^{\mathrm{X2C}}+H_{\mathrm{so}}^{\mathrm{DKH}},\label{approximateH}\\
H_{\mathrm{sf}}^{\mathrm{X2C}}&=&[\mathbf{h}_{+,\mathrm{sf}}^{\mathrm{X2C}}]_p^q \{a^p_q\}_F+\frac{1}{2} g_{pq}^{rs} \{a^{pq}_{rs}\}_F,\label{TCHsf}\\
H_{\mathrm{so}}^{\mathrm{DKH}}&=&[\mathbf{h}^{\mathrm{DKH}}_{\mathrm{so,1e}}+\mathbf{f}^{\mathrm{DKH}}_{\mathrm{so,2e}}]_p^q \{a^p_q\}_F, \label{SOCoper}
\end{eqnarray}
so as to treat SOC perturbatively when this is appropriate. Specific expressions for
the $\mathbf{h}_{+,\mathrm{sf}}^{\mathrm{X2C}}$, $\mathbf{h}^{\mathrm{DKH}}_{\mathrm{so,1e}}$, and $\mathbf{f}^{\mathrm{DKH}}_{\mathrm{so,2e}}$ integrals   can be found from Ref. \citenum{X2CSOC2}. $H_{\mathrm{sf}}^{\mathrm{X2C}}$ treats scalar relativity to infinite order
without computational overhead over the nonrelativistic counterpart. Moreover,
$H_{\mathrm{so}}^{\mathrm{DKH}}$ is variationally stable and is
computationally the same but more accurate than the Breit-Paul spin-orbit Hamiltonian.

At this stage, it can be stated that all relativistic Hamiltonians for all purposes of quantum chemical calculations
have been made available.  A complete and continuous `Hamiltonian Ladder' can hence be depicted\cite{IJQCrelH,PhysRep}, from which
one can just pick up a right Hamiltonian according to the target physics and accuracy.

\section{Electron Correlation}\label{SecEc}
\subsection{Conventional Methods}\label{SecGenerality}
In addition to PES, NES also contribute to electron correlation in the 4C relativistic framework.
Without the latter, the calculated correlation energy is always dependent on how the orbitals are generated, unlike
the FCI (full configuration interaction) solution of the Schr\"odinger equation. Yet, it should be stressed that
taking the NES as unoccupied (empty Dirac picture) is plainly wrong in this context, although it gives
rise to correct results for one-electron properties\cite{PCCPNES}. Only QED provides the correct
description of any two-body properties including correlation.
Given the huge gap (ca. $2c^2\sim1$ MeV) between the NES and PES,
a second-order treatment of the NES is sufficient. The MP2-like energy expression ($E^{(2)}_{+-}$)\cite{eQED} introduces
no significant computational overhead as compared with a higher-order treatment of correlation
within the manifold of PES. For the latter, any orbital product-based wave function methods can be used.
It is just that the loss of spin symmetry and concomitant complex algebra render the implementation more difficult.
Nevertheless, many sophisticated relativistic correlated wave function methods have been made available, including
4C or 2C many-body perturbation theory (MBPT)\cite{4C-MP21994,4C-CASPT22008,2C-MRPT22014,X2C-MRPT2-Li-2022}, coupled-cluster (CC)\cite{SOX2CAMF,4C-CCSD1995,4C-CCSD1996,2C-CC2001,2C-CC2005,2C-CC2007,
4C-IHFSCC2000,4C-IHFSCC2001,4C-FSCC2001,4C-CC2007,4C-CC2010,2C4C-CC2011,Lee-SOCC-2010,4C-CC2016,
2C-CC2017,Cheng2C-CC2018,Saue4C-CC2016,Saue4C-CC2018,2C-EOM-CCSD2019,Cheng2C-CCrev,SOC-CCSDT-GPU2021,CVS-EOM-CCSD2021},
configuration interaction (CI)\cite{4C-CISD1993,2C-CI1996,2C-CI2001,4C-GASCI2003,4C-CI2008,2C-CICC2012,Fleig2012,4C-icMRCI2015,2C-MRCI2020}, multiconfiguration self-consistent field (MCSCF)
\cite{RASSOC,CASSI1986,CASSI1989,4C-MCSCF1996,2C-CASSCF1996,2C-CASSCF2003,4C-CI-MCSCF2006,
2C-CASSCF2013,4C-MCSCF2008,4C-CASSCF2015,4C-CASSCF2018,LixiaosongX2CCASSCF},
density-matrix renormalization group (DMRG)\cite{4C-DMRG2014,MPSSI2021,X2C-DMRG-Li-2022,4C-DMRG2018,4C-DMRG-CC2020}, and FCI quantum Monte Carlo (QMC)\cite{4c-FCIQMC}. It should be clear from the outset that
such no-pair 4C and 2C methods are computationally identical after integral transformations.
Even the 4C integral transformations can be made identical with the 2C ones if the quasi-4-component (Q4C) relativistic Hamiltonian\cite{Q4C,Q4CX2C} is adopted. Therefore, it is merely a matter of taste to work with 4C or 2C approaches, for core properties of heavy elements
or valence properties involving $np$ ($n>5$) orbitals. For systems or properties where SOC is not very strong, it is more efficient
to postpone the treatment of SOC to the correlation step. Here, the gain in efficiency stems from the use of
real-valued orbitals and hence integrals, which allow for an easy incorporation of point group and full spin symmetries in constructing
the spin-free Hamiltonian matrix and both double group and time reversal symmetries in constructing
the SOC Hamiltonian matrix\cite{SpinUGA1,SOiCI}. For this reason, such approaches are usually called one-component (1C).
The interplay between SOC and electron correlation
can be accounted for in two ways\cite{Marian2001,marian2012SOCISC}, one-step (denoted as SOX for method X) or two-step
(denoted as XSO). The former type of approaches\cite{Lee-SOCC-2010,SpinUGA1,SOiCI,sf-X2C-EOM-SOC2017,WF-CC2008,WF-CC2011,DGCIa,DGCIb,DGCIc,DetSOCI1997,CI-SOC1998,SPOCK2006,HBCISOC2017,SOCASSCF2013}
aims to treat spin-orbit and electron-electron interactions on an equal footing, whereas the latter type of approaches
\cite{SOiCI,Hess1982,CIPSO1983,CI-SOC1997,SOCsingle-1,SOCsingle-2,SOCsingle-3,MRCI-SOC2000,
Teichteil2000,CASPT2-SOC2004,SI-SOC2006,EOMIPSO2008,mai2014perturbational,
DMRGSO2015,DMRGSO2016,nosiDMRGSO2016,cheng2018perturbative,ChengEOMCCSO2020,WangMP2020,Suo-SOC2021,vanWullen2021}
amounts to treating SOC after correlation, by
constructing and diagonalizing an effective spin-orbit Hamiltonian matrix over a small number of close-lying correlated scalar states.
While XSO's miss by construction spin-dependent orbital relaxations, SOX's can recover a large amount of such relaxations
(which is particularly true for SOCC\cite{Lee-SOCC-2010,sf-X2C-EOM-SOC2017}). Additional
spin-dependent orbital relaxation effects can be gained by further optimizing
the real-valued orbitals variationally in the presence of SOC, which is particularly beneficial to subsequent evaluation of response properties\cite{SOCASSCF2013}.

\subsection{Offbeat Methods}\label{SecSDS}
Given so many methods, a balanced and self-adaptive treatment of correlation and SOC
in strongly correlated systems containing heavy elements
is still highly desired, keeping in mind that configurations important for correlation and SOC resides in different regions of the Hilbert space.
To this end, we first recapitulate the SDS family of ``offbeat''\cite{Offbeat2015} methods for a balanced treatment of the static and dynamic components of correlation, including
SDSCI\cite{SDS}, XSDSCI\cite{SDS}, SDSPT2\cite{SDS,SDSPT2},
iCI\cite{iCI}, SiCI\cite{iCIPT2,iCIPT2New}, and iCIPT2\cite{iCIPT2,iCIPT2New} (see Table \ref{SDSmethods} for the acronyms).

\begin{center}
\small
\begin{threeparttable}
\caption{The SDS family of methods }
\begin{tabular}{lllcccccccc}
\toprule[0.5pt]
\multicolumn{1}{l}{acronym} & \multicolumn{1}{l}{full description} & reference \\
\midrule
SDSCI & static-dynamic-static configuration interaction&\cite{SDS}\\
SDSPT2&static-dynamic-static second-order perturbation theory&\cite{SDS,SDSPT2}\\
XSDSCI& extended SDSCI&\cite{SDS}\\
XSDSPT2&extended SDSPT2 &\cite{SDS,SDSPT2}\\
iCI   & iterative configuration interaction&\cite{iCI}\\
SiCI  & iCI with selection &\cite{iCIPT2,iCIPT2New}\\
iCIPT2& iCI with selection and PT2 &\cite{iCIPT2,iCIPT2New}\\
iCAS& imposed automatic selection and localization of complete active space &\cite{iCAS}\\
iCISCF(2)& iCI-based multiconfigurational self-consistent field theory  &\cite{iCISCF}\\
         &with inner-space PT2&  \\
SOiCI & one-component, one-step treatment of correlation and SOC & \cite{SOiCI}\\
iCISO & one-component, two-step treatment of correlation and SOC & \cite{SOiCI}\\
iVI   & iterative vector interaction for matrix diagonalization &\cite{iVI,iVI-TDDFT}\\
iOI      & iterative orbital interaction as a bottom-up solver of the SCF problem& \cite{iOI}\\
\bottomrule[0.5pt]
\end{tabular}
\label{SDSmethods}
\end{threeparttable}
\end{center}

The \emph{restricted} SDS framework\cite{SDS} for strongly correlated electrons was introduced heuristically as follows.
The $N_T$ exact (FCI) solutions for a given one-particle basis can generally be classified
into three portions, i.e., $N_P$ low-lying (primary), $N_S$ intermediate (secondary), and
$N_Q$ high-lying (external) states, depending on their mutual gaps $\Delta_{\mathrm{PS}}$ and $\Delta_{\mathrm{SQ}}$.
Although specific values for $N_P$, and $N_S$, and hence $N_Q$ ($=N_T-N_P-N_S$) are case dependent, it is always possible to
identify a decent gap $\Delta_{\mathrm{PS}}$ between the primary and secondary subspaces for a finite-sized molecular system, thereby leading to
 $1\le N_P\ll N_T$ (NB: the magnitude of $\Delta_{\mathrm{SQ}}$ and hence $N_S$ are irrelevant here). If  $N_P$ is just one,
the system is only weakly correlated and can well be described by a single-reference (SR) prescription.
Otherwise, a multireference (MR) treatment would be required to obtain the $N_P$ primary states simutaneously. The central questions are then (1)
how to construct many-electron functions to mimic such a general feature
(i.e., primary, secondary, and external subspaces) of the exact solutions and (2) how to
determine their weights in the $N_P$ primary states. For the former,
it is clear that only combinations of Slater determinants (SD) or configuration state functions (CSFs)
can have the right spectral feature. Based on these considerations, we write a FCI solution $|\Psi_I\rangle$ ($I\le N_P$) as
\begin{eqnarray}
|\Psi_I\rangle&=&\sum_{\mu=1}^{N_T} |\Phi_{\mu}\rangle C_{\mu I},\quad C_{\mu I}=\sum_{k=1}^{N_T} \tilde{C}_{\mu k}\bar{C}_{kI}\label{SVD}\\
&=&\sum_{k=1}^{N_T} |\tilde{\Phi}_k\rangle \bar{C}_{kI},\quad |\tilde{\Phi}_k\rangle=\sum_{\mu=1}^{N_T}|\Phi_{\mu}\rangle\tilde{C}_{\mu k}.\label{SVD2}\\
&\approx&\sum_{k=1}^{\tilde{N}}|\tilde{\Phi}_k\rangle \bar{C}_{kI}.\label{SVD3}
\end{eqnarray}
Here, $\{\Phi_{\mu}\}_{\mu=1}^{N_T}$ represent all possible $N$-electron SDs/CSFs,
while $\{\tilde{\Phi}\}_{k=1}^{N_T}$ are to be called contracted $N$-electron functions or simply ``states''.
It is trivial to see that $\{\tilde{\Phi}\}_{k=1}^{N_T}$ are just a nonsingular
transformation of $\{\Phi_{\mu}\}_{\mu=1}^{N_T}$ via $\tilde{\mathbf{C}}$, which can be viewed as
a step that renders the FCI Hamiltonian approximately diagonal (exactly diagonal if $\tilde{\mathbf{C}}=\mathbf{C}$!).
It is therefore clear that, as long as the contraction coefficients $\tilde{C}_{\mu k}$ are chosen properly, only a small number of states $\{\tilde{\Phi}_k\}_{k=1}^{\tilde{N}}$
are needed to represent accurately the FCI solutions $\{|\Psi_I\rangle\}_{I=1}^{N_P}$. To make this practical, the following questions
have to be addressed:
(a) What SDs/CSFs $\Phi_{\mu}$ are to be used as primitive $N$-electron basis functions;
(b) How their contraction coefficients $\tilde{C}_{\mu k}$ can be determined;
(c) How many states $\tilde{\Phi}_k$ are to be used;
(d) How the expansion coefficients $\bar{C}_{kI}$ are determined.
The procedure goes as follows.

A model space $P=\{\Phi_{\mu}; \mu=1,\cdots,d_R\}$ is first constructed (following, e.g., a selection procedure\cite{CIPSIb,iCIPT2New}),
which need not be a complete active space (CAS) nor that used to determine the one-particle orbitals.
The lowest $N_P$ solutions of the projected Hamiltonian $PHP$,
i.e., $\Psi^{(0)}_k=\sum_{\mu=1}^{d_R}\Phi_{\mu}\tilde{C}^{(0)}_{\mu k}$, which provide
either semi-quantitatively or qualitatively correct descriptions
of the $N_P$ exact states, are to be taken as the $N_P$ primary states. For convenience, we introduce the following projectors
\begin{eqnarray}
P_m&=&\sum_{k=1}^{N_P}|\Psi^{(0)}_k\rangle\langle\Psi_k^{(0)}|,\label{Pmdef}\\
P_s&=&P-P_m,\quad P= \sum_{\mu=1}^{d_R}|\Phi_{\mu}\rangle\langle\Phi_{\mu}|, \label{Psdef}
\end{eqnarray}
where $P_m$ and $P_s$ characterize the primary and secondary parts of the model space $P$, respectively.
As known from MBPT, the state-specific first-order corrections to $\{\Psi^{(0)}_k\}_{k=1}^{N_P}$, viz.,
\begin{eqnarray}
|\Xi_k^{(1)}\rangle&=&Q\frac{1}{E_k^{(0)}-H_0}QH|\Psi^{(0)}_k\rangle=\sum_{q\in Q}|\Phi_q\rangle \tilde{C}^{(1)}_{qk},\quad k\in[1, N_P],\label{PTWF1def}\\
Q&=&1-P, \label{Q1def}
\end{eqnarray}
are very effective in describing dynamic correlation and hence good candidates for the second (external) $N_P$ functions $\tilde{\Phi}_k$ in Eq. \eqref{SVD3}.
To account for changes in the static correlation (described by $\Psi_k^{(0)}$) due to the inclusion of dynamic
correlation (described by $\Xi^{(1)}_k$), the following \emph{not-energy-biased} Lanczos type of functions
\begin{eqnarray}
|\Theta_k\rangle&=&P_s H |\Xi^{(1)}_k\rangle\label{SecondaryWFdef}\\
&\approx&P_s^\prime H |\Xi^{(1)}_k\rangle,\quad P_s^\prime=\sum_{l=N_P+1}^{N_P+M_P}|\Psi_l^{(0)}\rangle\langle\Psi_l^{(0)}|\label{SecondaryWFdefapp}
\end{eqnarray}
can be introduced\cite{SDS} to mimic the third (secondary) $N_P$ functions $\tilde{\Phi}_k$ in Eq. \eqref{SVD3}.
Therefore, as a first attempt, Eq. \eqref{SVD3} reads
\begin{eqnarray}
|\Psi_I\rangle&=&\sum_{k=1}^{N_P}|\Psi_k^{(0)}\rangle\bar{C}_{kI}+\sum_{k=1}^{N_P}|\Xi_k^{(1)}\rangle\bar{C}_{(k+N_P)I}
+\sum_{k=1}^{N_P}|\Theta_k\rangle\bar{C}_{(k+2N_P)I}.\label{ixc1}
\end{eqnarray}
The yet unknown coefficients $\bar{\mathbf{C}}$ are to be determined by the generalized secular equation
\begin{eqnarray}
\bar{\mathbf{H}}\bar{\mathbf{C}}=\bar{\mathbf{S}}\bar{\mathbf{C}}\mathbf{E}, \label{EigenEQ}
\end{eqnarray}
where the reduced Hamiltonian $\bar{\mathbf{H}}$ and metric $\bar{\mathbf{S}}$ have the following structure
\begin{eqnarray}
\bar{\mathbf{H}}&=&\begin{pmatrix}
P_mHP_m& P_mHQ& P_mHP_s\\
QHP_m&QHQ&QHP_s\\
P_sHP_m&P_sHQ&P_sHP_s\end{pmatrix}_{3N_p\times 3N_p} \label{Hblcok}\\
&=&\begin{pmatrix}E_{k}^{(0)}\delta_{kl} & \langle\Psi_k^{(0)}|H|\Xi_l^{(1)}\rangle &0 \\
 \langle\Xi_l^{(1)}|H|\Psi_k^{(0)}\rangle& \langle\Xi_k^{(1)}|H|\Xi_l^{(1)}\rangle &\langle\Xi_k^{(1)}|H|\Theta_l\rangle\\
 0&\langle\Theta_l|H|\Xi_k^{(1)}\rangle& \langle\Theta_k|H|\Theta_l\rangle\end{pmatrix}, \quad k, l \in[1, N_P],\label{Hmat}\\
\bar{\mathbf{S}}&=&\begin{pmatrix}\delta_{kl} & 0 &0 \\
 0& \langle\Xi_k^{(1)}|\Xi_l^{(1)}\rangle & 0\\
 0&0& \langle\Theta_k|\Theta_l\rangle\end{pmatrix}, \quad k, l \in[1, N_P].
\end{eqnarray}
It has been shown\cite{SDS} that the weights of $\Psi_k^{(0)}$, $\Xi_k^{(1)}$, and $\Theta_k$ in $|\Psi_I\rangle$ \eqref{ixc1}
do decrease, thereby justifying the ``static-dynamic-static'' characterization. As a matter of fact, the secondary role
of $\{|\Theta_k\rangle\}_{k=1}^{N_P}$ is obvious from the structure of $\bar{\mathbf{H}}$ \eqref{Hmat}: They interact with
$\{|\Psi^{(0)}_k\rangle\}_{k=1}^{N_P}$ only indirectly, through interactions with $\{|\Xi^{(1)}_k\rangle\}_{k=1}^{N_P}$.
 Note in passing that $P_s^\prime$ \eqref{SecondaryWFdefapp}
with $N_P\le M_P\ll d_R-N_P$
is a very good approximation to the full $P_s$ \eqref{Psdef} but renders the evaluation of the matrix elements much easier.
Except for the particular form \eqref{SecondaryWFdef},
other types of secondary states are also possible. For instance, the second set of $N_P$ states
$\{\Psi_{k+N_P}^{(0)}\}_{ k=1}^{N_P}$ of the projected Hamiltonian $PHP$ are also good candidates for $\{\Theta_k\}$ in view of intermediate Hamiltonian theory\cite{IHb}
and are computationally very efficient.

Several remarks can be made here.
\begin{enumerate}[(1)]
\item The three sets of functions, $\{|\Psi^{(0)}_k\rangle\}_{k=1}^{N_P}$, $\{|\Xi^{(1)}_k\rangle\}_{k=1}^{N_P}$, and $\{|\Theta_k\rangle\}_{k=1}^{N_P}$, are treated independently, so that the dimension of Eq. \eqref{EigenEQ} is just three times the number ($N_P$) of wanted states,
\emph{irrespective} of the numbers of correlated electrons and orbitals. This Ansatz is best characterized as ``internally and externally contracted minimal MRCI with
singles and doubles and augmented with secondary states'' (ixc-MRCISD+s), denoted as SDSCI for short\cite{SDSRev}.
The diagonalization of large matrices only for a smaller number ($N_P$) of states
in \emph{unrestricted} SDS methods (e.g., internally contracted multireference configuration interaction
with singles and doubles (ic-MRCISD)\cite{icMRCI3a,icMRCI3b}) is completely avoided here. It is therefore clear that
the computational cost of SDSCI is very much the same as that MRPT3. Being variational,
SDSCI differs fundamentally from other three-subspace-based methods\cite{CIPSIb,Selection3},
which are state-specific MRPT2's belonging to the ``static-then-dynamic'' family\cite{iCI}. Instead,
SDSCI is similar to superdirect CI (Sup-CI)\cite{FAST-CIa,Sup-CIa,Sup-CIb} in that only a small number of contracted $N$-electron
functions are used to project the Hamiltonian, so as to avoid the explicit generation of long CI eigenvectors. However, instead of the
(restricted) SDS philosophy,
Sup-CI tries to mimic the performances of various iterative diagonalization approaches, by taking $\{\Psi_k^{(0)}\}$ (or $\{\Phi_{\mu}\}$) and
different variants of first-order corrections to them as independent $N$-electron basis functions\cite{Sup-CIb}.
That is, SDSCI and Sup-CI become identical only for the simplest case: one state spanned by two functions $\Psi_0^{(0)}$ and $\Xi_0^{(1)}$.
%

\item SDSCI can be reduced to multi-state (MS) MRPT2
by setting $\bar{C}_{kI}=\bar{C}_{(k+N_P)I}$ and $\bar{C}_{(k+2N_P)I}=0$ in Eq. \eqref{ixc1}, leading to
 \begin{eqnarray}
 \Psi_I^{(1)}=\sum_{k=1}^{N_P}(\Psi_k^{(0)} + \Xi_k^{(1)}) \bar{C}_{kI}.\label{PTWF}
\end{eqnarray}
The Hamiltonian matrix \eqref{EigenEQ} correct to second order is then of dimension $N_P$, viz.,
 \begin{eqnarray}
 (\bar{H}^{[2]})_{kl}&=&\frac{1}{2} \{\langle \Psi_k^{(0)} + \Xi_k^{(1)}|H|\Psi_l^{(0)}\rangle  + \langle \Psi_k^{(0)}|H|\Psi_l^{(0)}+\Xi_l^{(1)}\rangle\}\nonumber\\
 &=&\langle \Psi_k^{(0)}|H|\Psi_l^{(0)}\rangle
 +\frac{1}{2}\{\langle\Xi_k^{(1)}|H|\Psi_l^{(0)}\rangle +\langle\Psi_k^{(0)}|H|\Xi_l^{(1)}\rangle\}.\label{MSMRPT2}
\end{eqnarray}
It is well known that such MS-MRPT2's, albeit following the ``diagonalize-perturb-diagonalize'' procedure,
do not have the efficacy of relaxing sufficiently the coefficients
of the model functions $\{\Phi_{\mu}\}_{\mu=1}^{d_R}$ (i.e.,
$\sum_{k=1}^{N_P}\tilde{C}^{(0)}_{\mu k}\bar{C}_{kI}$ are not much different from $\tilde{C}_{\mu k}^{(0)}$)
and therefore belong also to the ``static-then-dynamic'' category of methods\cite{iCI}. To stay within
the SDS framework, a CI-like MRPT2, SDSPT2, has been introduced\cite{SDS,SDSPT2}, which amounts to
replacing the $QHQ$ block in Eq. \eqref{Hblcok} with $QH_0Q$
before the diagonalization step is taken. Unlike MS-MRPT2, which treats single and multiple states differently,
SDSPT2 does this in the same way. More interestingly,
all MS-MRPT2's based on Eq. \eqref{MSMRPT2} can be obtained from SDSPT2 for free, as long as the same zero-order Hamiltonian $H_0$ and perturbers are used. This is because the former amount to just picking up the $P_mHP_m$ and $P_mHQ$ blocks of $\bar{\mathbf{H}}$ \eqref{Hblcok}.
SDSPT2 combines the good of standard MBPT (with one-to-one fixed combination of $\Psi_k^{(0)}$ and $\Xi^{(1)}_k$), intermediate Hamiltonian
(with buffer to avoid intruder states and meanwhile bring in secondary static correlation),
and straight CI (diagonalization) and is hence ``offbeat''\cite{Offbeat2015}.

\item SDSCI can also be extended\cite{SDS} in a simple way (denoted as XSDSCI hereafter): Those external functions with coefficients $\tilde{C}_{qk}^{(1)}$
larger than the preset threshold $Q_{\mathrm{min}}$ can be treated as independent functions, so as to increase the variational degrees of freedom.
Alternatively, the extension of SDSCI can be achieved by decomposing the
virtual orbitals into two disjoint sets via the virtual space decomposition (VSD)
approach\cite{SDSRev,SOiCI}, with excitations to the low-lying set of virtual orbitals treated as independent functions.
In any case, the dimension of XSDSCI is kept small (to the extent that the diagonalization can be performed trivially).
If the $QHQ$ block of XSDSCI is replaced with with $QH_0Q$ before diagonalization, we would obtain XSDSPT2 as an extension of SDSPT2.
In short, SDSPT2, SDSCI, XSDSPT2, and XSDSCI approach steadily to ic-MRCISD, for they all share the same first-order
interacting space and differ only in how the expansion coefficients are determined.
In particular, XSDSPT2/SDSCI (SDSPT2) can be obtained from XSDSCI (SDSCI) for free, whereas XSDSCI (or SDSCI) can be taken as initial guess for
solving iteratively ic-MRCISD. It can be sure that the internal consistency among SDSPT2, SDSCI, XSDSPT2, XSDSCI, and (the first few
iterations of) ic-MRCISD is a strong indicator for highly accurate results.

\item Like ic-MRCISD, SDSPT2, SDSCI, XSDSPT2, and XSDSCI are not size consistent. However, the errors
can readily be cured\cite{SDSRev} by using the Pople correction\cite{Pople1977}, which is particularly suited to internally contracted
Ans\"atze. Extensive tests have revealed that SDSPT2 with the Dyall Hamiltonian\cite{Dyall-H} as $H_0$
performs virtually the same as
the well-established MS-NEVPT2\cite{MSNEVPT2} for low-lying states of not-strongly correlated systems\cite{SDSRev},
but clearly better than the latter for cases with multiple nearly degenerate states\cite{SDSPT2}. On the other hand,
SDSCI usually differs only marginally from ic-MRCISD, needless to say XSDSPT2 and XSDSCI.

\item Another extension of SDSCI is to take its
the eigenvectors (cf. Eq. \eqref{EigenEQ}) as new primary states $\{|\Psi_k^{(0)}\rangle\}_{k=1}^{N_P}$,
such that the SDS procedure can be iterated until convergence.
Such iterative configuration interaction (iCI)\cite{iCI} accesses in each iteration (defined as macroiteration)
a space that is higher by two ranks than that of the preceding iteration.
Up to $2i$-tuple excitations (relative to the initial primary
space) can be accessed if $i$ macroiterations are carried out.
A few microiterations $j$ can be invoked in each macroiteration so as to relax the contraction coefficients
(because of this, the quality of $H_0$ in Eq. \eqref{PTWF1def} is immaterial).
As such, every iCI($i, j$) corresponds to a physically meaningful model. For instance, iCI(2,2) starting with CASSCF
(complete active space (CAS) self-consistent field)\cite{CASSCF}
is approximately uncontracted MRCISDTQ and produces a lower energy for equilibrium water than DMRG (cf. Table I in Ref. \citenum{iCI}).
It has been shown both theoretically and numerically that iCI can converge quickly from above to FCI even when starting with a very poor initial guess.
More generally, iCI can be viewed as a particular sequential, exact partial diagonalization of a huge matrix, by getting first the roots of one portion
of the matrix and then those of an enlarged portion, until the full matrix has been sampled,
whereas its microiterations can be generalized to an iterative vector interaction (iVI) approach\cite{iVI}
for arbitrary roots of a given matrix treated as a whole.
In particular, by combining with the energy-vector following technique,
iVI can directly access interior roots belonging to a predefined window, without knowing the number and characters of the roots\cite{iVI-TDDFT,AutoPST2022}.
Since the Hartree-Fock/Kohn-Sham equation can be viewed as a one-electron FCI problem, the idea of iCI can also be employed as a bottom-up solver
of such mean-field equations, leading to iOI (iterative orbital interaction)\cite{iOI}.
\item As an exact FCI solver, the computational cost of iCI\cite{iCI} stems solely from the construction of the Hamiltonian matrix $\mathbf{H}$ in
the basis of SDs/CSFs before contracted to
$\bar{\mathbf{H}}$ \eqref{Hmat} for diagonalization. Given the sparsity of CI vectors, it is natural to combine iCI
with configuration selection (i.e., SiCI) so as to construct a variational space $P_m$ that is as compact as possible.
The dynamic correlation from the complementary space $Q=1-P_m$ can then be accounted for by PT2.
The resulting iCIPT2 has been described in great detail before\cite{iCIPT2,iCIPT2New} and is hence not repeated here.
It suffices to say that iCIPT2 is a near-exact approach\cite{Blindtest} for arbitrary open-shell systems thanks to the use of CSFs as
the many-electron basis, and rivals other similar approaches in efficiency due to the use of tabulated unitary group approach (TUGA)\cite{iCIPT2}
for the evaluation and reusage of the basic coupling coefficients between unstructured CSFs.
Moreover, when combined with the iVI approach\cite{iVI}, iCIPT2 can
access directly excited states that have no overlap with the low-lying ones.
\item SiCI\cite{iCIPT2,iCIPT2New} can be combined with CASSCF (denoted as iCISCF\cite{iCISCF}), so as to achieve large active space
calculations (e.g., 60 electrons in 60 orbitals). If a PT2 correction is further carried out within the $P_s$ space (left
over by selection within the CAS $P$), the resulting iCISCF(2) is essentially the same as CASSCF in energy.
Such large active space calculations can further be facilitated by iCAS\cite{iCAS} for imposed selection and localization of CAS. Specifically,
iCAS starts with a set of valence atomic orbitals (VAO) or primitive fragment localized molecular orbitals (pFLMO) from subsystem calculations\cite{FLMO1,Triad,ACR-FLMO,FLMO3}. Such VAOs/pFLMOs
are then converted to an equivalent set of pre-LMOs of occupation numbers 2, 1 or 0,
to select (by matching) precisely the same number of doubly, singly, and zero occupied ROHF (restricted open-shell Hartree-Fock) LMOs\cite{FLMO1,Triad,ACR-FLMO,FLMO3} as initial guess. If wanted, the number of active
orbitals can be reduced by first semicanonicalizing those pre-LMOs of a fragment (e.g., a non-broken aromatic ring) and then
picking up those of right ``energies'' (NB: semicanonicalization
means here that the doubly, singly, and zero occupied pre-LMOs are canonicalized separately). Since such regional pre-CMOs
are still localized on the chosen fragment, they can be
used simply as pre-LMOs, provided that the molecular ROHF-LMOs
localized on the same fragment are also semicanonicalized in the same way.
Having determined the guess orbitals, the localized CASSCF orbitals of two adjacent iterations
are further enforced to match most, so as to guarantee the same
core, active, and virtual subspaces throughout the iterations.
Since the VAOs/pFLMOs and hence pre-LMOs do not change with geometric displacements,
iCAS can always select the same guess orbitals for the whole potential energy surface (PES).
On the other hand, the orbital matching procedure helps to reduce the possibility for
 the orbital optimization to converge to different CASSCF solutions at different geometries.
Numerous examples have shown\cite{iCAS} that iCAS can indeed produce smooth PESs, even in very difficult situations.
As such, issues pertinent to CASSCF, e.g., the choice of which and how many orbitals as active orbitals, the swap of orbitals during
SCF iterations, the difficulty of maintaining the same CAS in scanning PESs, and the exponential scaling of CASSCF,
are all minimized to a large extent. Not only so, the automatically generated local CASSCF orbitals
are very useful in elucidating the mechanisms of complex chemical reactions\cite{Polymerization2022}.
\end{enumerate}

\begin{figure}[H]
	\centering
	\includegraphics[width=0.8\textwidth]{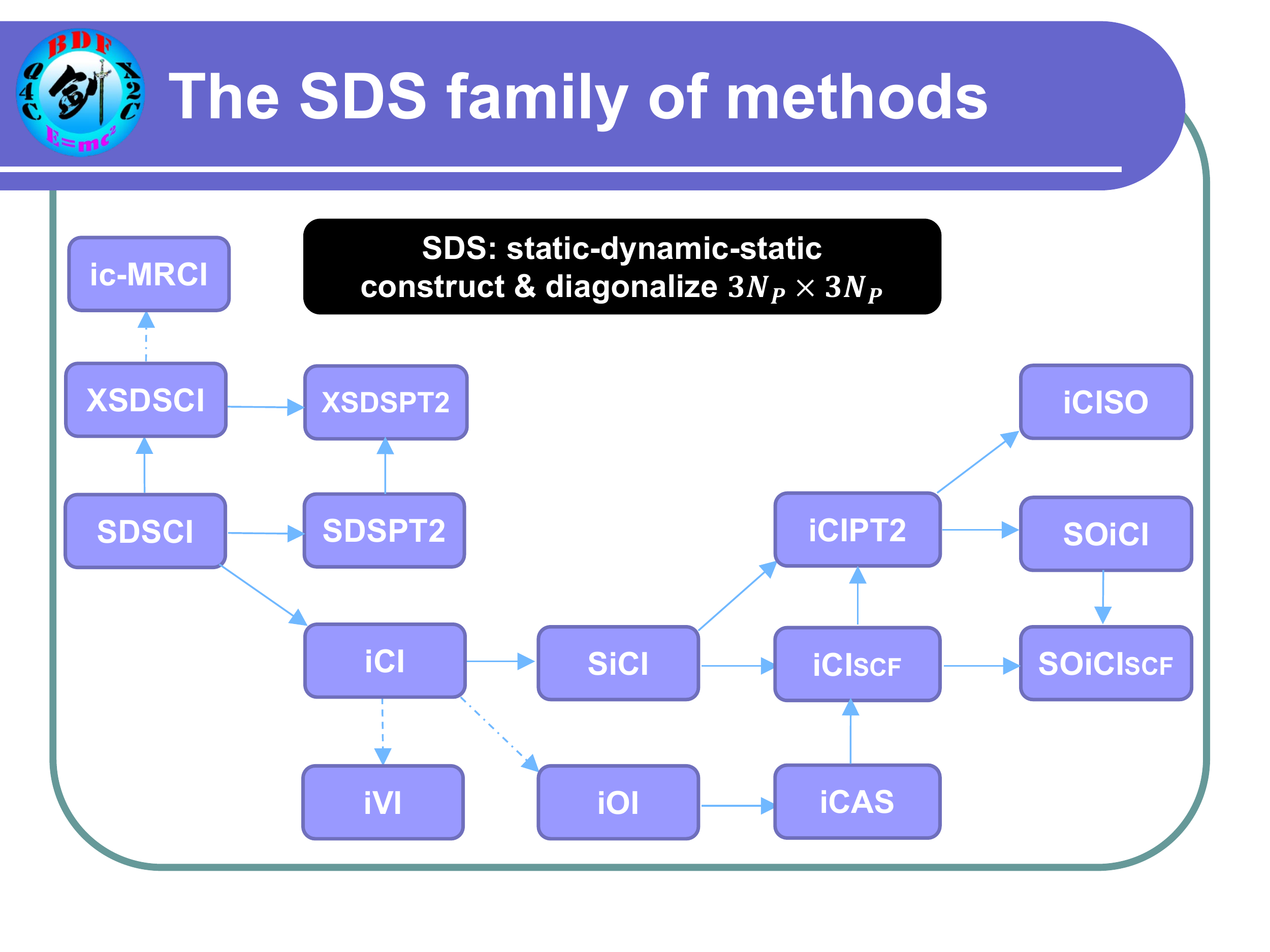}
	\caption{The SDS family of methods (cf. Table \ref{SDSmethods})}
	\label{SDSfigure}
\end{figure}

Having summarized the SDS family of methods for correlation (see Fig. \ref{SDSfigure} for their
mutual relations), we are now readily to address the issue of how to handle correlation and SOC in a balanced manner. At first glance, this can be achieved by means of selection of configurations important for either correlation or SOC,
such that their union can provide a good description of both correlation and SOC. However, there exists a caveat here:
There may exist some orbitals (e.g., those high in energy but local in space) with appreciable SOC matrix elements but
their mirror occupied orbitals are not in the correlation space due to the limited number of correlated electrons.
In this case, a balanced description of correlation and SOC cannot be achieved at all\cite{SOiCI}. So is
the opposite situation (i.e., the low-lying occupied orbitals are correlated but their mirror virtual orbitals are not).
To avoid such situations, the VSD approach\cite{SDSRev,SOiCI} can be employed to map out a set of occupied and virtual orbitals
from the all-electron calculation, via a small set of orthonormal AOs derived by
symmetric orthonormalization of, e.g., a SRECP (scalar relativistic effective core potential) DZP basis.
The so-selected orbitals along with the valence electrons span a sufficiently large active space within which
the selection of configurations can be performed for correlation and SOC.
It has been demonstrated\cite{SOiCI} that this idea does work very well. The use of localized HF/CASSCF orbitals\cite{FLMO1,Triad,ACR-FLMO,FLMO3,iCAS}
is even better, thanks to the short-range nature of SOC. Since SOC is strongly dominated by the one-body terms,
a second-order perturbative treatment of dynamic correlation, on top of
a well-controlled variational space, should be sufficient.

\section{Perspectives}\label{Conclusion}
Having summarized the relativistic Hamiltonians and wave function-based correlation methods, we finally outline a road map
for solving the titled ``equation''. It is obvious that the very first step is to treat
the VP-ESE \eqref{QDef} (fitted into a model operator\cite{ShabaevModelSE,ShabaevModelSEcode2018})
variationally at the mean-field level via, e.g., eQED@X2C-iCISCF\cite{iCISCF}, so as to obtain the \emph{genuine} QED effects (Lamb shifts)
as well as screenings of them. Those ``derived'' QED effects (photon self-energies and vertex corrections or
equivalently virtual-pair effects)\cite{eQED} can then readily be accounted for by the $E^{(2)}_{+-}$ component of
$E^{(2)}_{\mathrm{eQED}}$ (cf. Fig. \ref{FigureE2}). It is just that the frequency-dependent Breit interaction
must be invoked here. Otherwise, $E^{(2)}_{+-}$ would be severely overestimated. What is left is then
the correlation within the manifold of PES. Before a highly efficient implementation of eQED@X2C-iCIPT2\cite{iCIPT2,iCIPT2New} is realized,
the double group and time reversal adappted 1C-SOiCI approach\cite{SOiCI}, which treats correlation and SOC on an equal footing for arbitrary open-shell systems, can be adopted. Spin-dependent orbital relaxation effects can further be gained by optimizing the scalar orbitals
variationally, although it is usually not necessary because of the use of large enough active spaces.
Of course, all these are personal points of view. Other no-pair relativistic correlated
approaches, e.g.,
4C/X2C-MRCC/DMRG/FCIQMC\cite{4C-DMRG2018,4C-DMRG-CC2020,4c-FCIQMC}, are highly promising as well.

\section*{Acknowledgment}
This work was supported by the National Natural Science Foundation of China (Grant Nos. 21833001 and 21973054),
Mountain Tai Climbing Program of Shandong Province, and Key-Area Research and Development Program of Guangdong Province (Grant No. 2020B0101350001).

\section*{CONFLICT OF INTEREST}
The author declares no conflict of interest.

\section*{Data Availability Statement}
The data that supports the findings of this study is available within the article.

\section*{ORCID}

\url{https://orcid.org/0000-0002-1630-3466}

\clearpage
\newpage

\bibliography{BDFlib}


\end{document}